\documentclass[
 aps, pra,
 amsmath,amssymb,
 12pt,
 final,
tightenlines,
 nofootinbib,
 superscriptaddress,
 ]
{revtex4}

\usepackage[utf8x]{inputenc}
\UseRawInputEncoding
\usepackage[english]{babel}
\usepackage{graphicx}

\def\degr{\hbox{$^\circ$}}
\begin{document}

\title{Instabilities of the Kinematic State of the Atmospheres of Single C-Rich Post-AGB Stars}
\author{Valentina G.~Klochkova}
\email{Valentina.R11@yandex.ru}
\author{Vladimir~E.~Panchuk}
\author{Nonna~S.~Tavolzhanskaya}
\author{Maksim~V.~Yushkin}

\affiliation{Special  Astrophysical Observatory,  Nizhnij Arkhyz, 369167 Russia}

\sloppypar
\vspace{2mm}
\noindent

\begin{abstract}
To search for and study the instabilities in the atmospheres of selected post-AGB stars,
we have performed a long-term high-resolution spectroscopy (R$\ge$60\,000) with the spectrograph NES
of the 6-meter BTA telescope. Low-amplitude pulsations, splitting and/or asymmetry of the
absorption profiles with a low excitation potential (mainly absorptions of s-process metals),
as well as variability of a complex H$\alpha$ profile have been registered in the optical
spectra of single stars associated with the IR sources IRAS\,z02229+6208, IRAS\,04296+3429,
IRAS\,07134+1005, IRAS\,07430+1115, IRAS\,19500$-$1709, IRAS\,22223+4327, and IRAS\,23304+6147
that had previously undergone the 3-d dredge-up. The maximum pulsation amplitude A$_{\rm Vr}$
was detected for the stars in the IRAS\,07134+1005 and IRAS\,19500$-$1709 systems, which have
the maximum temperatures among the stars studied. Stratification of radial velocity in the
atmosphere was found for two stars in the sample. The luminosity of the studied stars was
estimated based on the intensity of the IR oxygen triplet O\,I(7774).
Moreover, a luminosity of log${\rm (L/L_{\odot})}\approx$3.1 was obtained for the star in the
IRAS\,07430+1115 system within the typical values for post-AGB stars luminosity, which eliminates
the paradox of the luminosity and the initial mass of this object. 
\end{abstract}

\maketitle

\section{Introduction}

The objects of this study are single stars after the Asymptotic Giant Branch (AGB) with  initial
masses in the range of about $1\div8\mathcal{M}_{\odot}$, which have undergone the evolutionary
phases with nucleosynthesis and the third mixing, as well as episodes with mass loss due to
stellar wind at differents pace. During the hydrogen and helium layer burning stages,
stars experience mass loss at the rates in the range of 10$^{-8}\div10^{-4}$$\mathcal{M}_{\odot}/$yr
 [1].  The rate of mass loss due to the wind decreases significantly during the transition from the AGB
 stage to the subsequent post-AGB stage (see papers [2, 3]).
This evolutionary transition, taking place, according to the theoretical calculations of Miller
Bertolami [4], over a period of time from several hundred to many thousands of years
for stars of different masses, comes along with the separation of the envelope from the star.
In essence, this short-lived post-AGB stage of the evolution of intermediate-mass stars  is
a fast transition from a star to a planetary nebula, which is why they are often called protoplanetary
nebulae (PPN).

The sample of thoroughly studied stars with overabundances of carbon and heavy metals (hereinafter,
C-rich post-AGB stars) is currently small, since their apparent brightness is significantly reduced
due to the extinction by their own dust envelopes. However, the interest of astrophysicists in AGB
stars and their closest descendants, post-AGB stars is steadily growing. This growth is due, first
of all, to their influence on the evolution of the chemical composition of galaxies, since in the
depths of these stars, being at a short-term evolutionary stage and possessing a complex internal
structure, brightness and spectrum variabilities, physical conditions arise for the synthesis of
nuclei of heavy metals and their subsequent dredge-up into the stellar atmosphere. Due to these
processes, AGB stars are the main suppliers (over 50\%) of all elements heavier than iron, synthesized
owing to the s-process, the essence of which is a slow neutronization of nuclei. The results of
calculations of the synthesis and ьшчштп of elements are presented by Herwig [5], Di Criscienzo et
al. [6], Liu et al. [7].

According to current concepts of the evolution of AGB stars [5], after depletion
of helium in the core, the structure of the star changes  -- a degenerate carbon-oxygen core is
formed, surrounded by alternative energetically active layers of helium and hydrogen combustion.
Most of the time, the energy release is provided by the hydrogen layer, while the helium layer
adjacent to the degenerate core remains inert. Between these energy-releasing layers, a thin
layer is formed, the so-called ``He-intershell'', where, as the temperature increases, helium
combustion and carbon accumulation occur, thus creating conditions for the most important reaction,
$^{13}$C${\alpha}$,n)$^{16}$O. Owing to this reaction neutrons are formed, providing a subsequent
synthesis of heavy metal nuclei and effective mixing (see a detailed description of these
processes and the necessary references in  Cristallo et al. [8]. The dredge-up of freshly
produced atoms into the circumstellar environment is caused by the instability in the atmospheres
of supergiants due to the stellar wind and envelope ejection, pulsations and other kinematic
processes,  which is also manifested in the features of their spectra.

Photometric and spectral variability of post-AGB stars has been actively studied for
several decades after their identification with IR sources. Note the results obtained over several
decades by B.~Hrivnak’s group within the program  of searching and refining the periods of brightness
variability for a dozen C-rich post-AGB stars.
In their recent publication, Hrivnak et al. [9], combining the long-term photometric
data from the Valparaiso University Observatory (VUO, IN 46383, USA) with the available ASAS data,
determined  the amplitude of brightness variability $\Delta$V and  a set  of periods for them. These
authors emphasized that the complex nature of the light
curve and the period change of post-AGB stars can be caused by several physical processes:
pulsations in the atmosphere, non-sphericity of the envelope, and non-uniformity of the surface
layer of the star due to the presence of convective spots. For a sample of far evolved stars
the members of V.P.~Arkhipova’s team from  the SAI MSU have carried out multicolor photometry
and spectral  observations with a moderate spectral resolution. These data allowed us to study
the brightness variability, color excesses, estimate  the variability periods and identify
the main features of the spectra of a set of stars on the evolutionary path from the AGB
stage to planetary nebulae (see in particular the papers by Arkhipova et  al. [10--13].
The papers published over three decades with the results of the Arkhipova et al. team often
served  as the basis for the subsequent high-resolution  spectroscopy of a number of stars
at the BTA [14--16].

At the end of the IRAS mission, many IR sources were identified with far evolved AGB and
post-AGB stars, which served as the basis for a detailed  spectroscopy program of a sample
of these objects  at the 6-meter BTA telescope. The analysis of our  spectral data was aimed
primarily at determining  the fundamental parameters and features of the atmospheric chemical
composition  of the program stars. Subsequently, spectral monitoring was performed for the
selected program stars with the most intriguing spectral features to search for a
variability of specific spectral features and the radial velocity pattern. However, stars
with large IR flux excesses are usually faint in the visible range due to a significant
circumstellar extinction. Only a small sample of faint in the visible range central
stars of  IR sources are accessible at large telescopes for the optical spectroscopy
with high spectral resolution. In our survey, we will consider the observed features for
the following objects: IRAS\,z02229+6208 (further referred to as IRAS\,z02229), IRAS\,04296+3429
(IRAS\,04296), IRAS\,07134+1005 (IRAS\,07134), IRAS\,07430+1115 (IRAS\,07430),
IRAS\,19500$-$1709 (IRAS\,19500), IRAS\,22223+4327 (IRAS\,22223) and IRAS\,23304+6147 (IRAS\,23304).

Numerous results that we have obtained during the spectroscopy and spectral  monitoring
are summarized in the following surveys: Klochkova [17--19]  and  Klochkova et al. [20].
The paper by Klochkova [21] presents some
parameters of circumstellar envelopes that we have determined, often non-spherical
and structured, incombination with radio spectroscopy data. As our program progressed,
it was necessary to determine or clarify the evolutionary status of a number of objects
under study. An excellent example is the results obtained for the supergiant V1302\,Aql
in the system of the IR source IRC+10420. For a long time, this star was classified
as a low-mass supergiant at the post-AGB stage, but the subsequent refinements
of its parameters (luminosity and features of the atmospheric chemical composition
by Jones et al. [22]; Oudmaijer et al. [23]; Klochkova et al. [24]
allowed us to classify V1302\,Aql as one of the most massive evolved stars at the
yellow hypergiant stage, near the high-temperature boundary of the Yellow Void [25].
Note the extremely high luminosity of V1302\,Aql  obtained by Oudmaijer et al. [26]
based on the distances from the Gaia~DR3 data [27].
In the list of numerous post-AGB stars given by Oudmaijer et  al. [26], the
luminosity of V1302\,Aql is at least
2--3 orders of magnitude higher than that of other G--supergiants. The status of
the star BD$-11\degr$1178 in  the IRAS\,05238$-$0626 system was changed just as
significantly as well. High-resolution spectral monitoring revealed the star’s
binarity (SB2), but the analysis of the data set did not confirm that the star
belongs to the post-AGB stage [28]. As a result, the authors  [28]  concluded that
BD$-11\degr$1178 is a young pair of pre-main sequence  F-stars and suggested its
membership in the 1c subgroup  of the Ori OB1 association.

The main objective of this survey is to analyze the temporal behavior of spectral
features and radial velocity patterns for a sample of single C-rich stars,
members of a homogeneous subgroup of post-AGB stars with a detached envelope,
a double-humped energy distribution (SED) in the spectrum, and atmospheres
enriched in heavy metals of the s-process. It is obvious that to search for the
temporal variations in the spectrum and the velocity field for each object
of the program, it is necessary to perform multiple observations, spaced by dates.
Here we summarize the results obtained and published for individual stars earlier.
The results of spectroscopy of C-rich stars in the IRAS\,04296, IRAS\,07134, IRAS\,19500,
IRAS\,22223 and IRAS\,23304 source systems are published in the following papers:
[29 -- 33], respectively. Later, some additional results of a more detailed
study were published using the new  high-quality spectra of C-rich stars in the
IRAS\,07134 [34]  and IRAS\,23304 [35]  systems.

Here, along with these more studied objects, we present here in more detail
the results of monitoring of a poorly studied star in the IRAS\,07430 system. A
significant part of this survey is devoted to comparing the features of this
star with its kindred star in the IR-source IRAS\,z02229. The similarity of this
pair of stars was previously declared by Reddy et al.[36]  on the basis
of similar values of their fundamental parameters, metallicity, and features of
the chemical composition of their atmospheres. Having analyzed the high-resolution
echelle spectroscopy data obtained with the McDonald Observatory 2.7-meter telescope
spectrograph, Reddy et al. [36] published the fundamental parameters and
detailed chemical compositions of the atmospheres of a pair of stars in the
IRAS\,z02229 and IRAS\,07430 systems. They concluded that the metallicity of the
stars is reduced: [Fe/H]=$-0.5$dex, given large excesses of carbon
[C/Fe]=$+0.8$dex and the s-process heavy metals [s/Fe]=$+1.4$dex. These chemical
abundances  indicate that both stars have passed the AGB evolutionary stage and
the third mixing. The available long-term photometric data series of IRAS\,z02229
and IRAS\,07430, together with the ASAS data [37], allowed to reveal the long-term
brightness variabilityof both stars with amplitudes of 0$^m$.70 and 0$^m$.29 [9].

The kinematic data for the C-rich post-AGB star associated with the IR source
IRAS\,z02229 were published by Klochkova and Panchuk [38]. Due to its
high color excess, this object belongs to the so-called ERO  C-rich stars,
congruent to the type of stars with forceful envelopes (Extremely Red Objects),
introduced by Groenewegen [39]. As follows from the data in Table~1, which
contains some basic data about the studied stars,   the total  extinction in the akin 
star in the system of the IRAS\,07430 source is significantly lower. This difference 
can be largely explained by the low extinction in the interstellar medium of the 
nearby object IRAS\,07430 (parallax $\pi\ge$3\,mas according to Gaia~DR3 [27], and
its location quite high above the Galactic plane. For this system, as well as for 
its kindred object IRAS\,z02229, there was no reiteration of a high-resolution
optical spectrum for a long time. The spectral monitoring of the star in the IRAS\,07430
system is complicated by its weak visual brightness (B=13.9$^m$, V=12.8$^m$).
These factors served as our incentive to observe the star with the spectrograph of
the 6-m telescope, which was performed on arbitrary dates over 2018$\div$2024.
In Section~2 we briefly describe the observations and data reduction methods.
In Section~3 we summarize our results and compare them with those published
earlier for the kindred objects, and in Section~4 we present a discussion of
the results and the main conclusions.

\section{Echelle spectroscopy at BTA}

The main set of spectra of C-rich stars during the program was obtained with
the NES echelle spectrograph [40], permanently mounted at the Nasmyth focus of
the 6-m BTA telescope. The spectral resolution of
the NES spectrograph is  $\lambda/\Delta\lambda\ge$60\,000, $\rm S/N$ varies
by about one and a half times along the echelle order in the spectra.
In addition, the signal is significantly reduced in the short-wave part of
the echelle frame due to a decrease in the flux from cold stars and
a significant extinction of the star radiation in its envelope. Currently,
the  NES spectrograph is equipped with a large-format CCD chip with 4608$\times$2048
elements sized 0.0135$\times$0.0135\,mm, the readout noise is 1.8e$^−$. The registered
spectral range in our spectra is usually $\Delta\lambda$=470--778\,nm. However,
the optical scheme of the NES allows changing the spectral range if necessary
(see the examples in the publications [41].
To reduce the light losses not falling of spectral resolution, the NES spectrograph
is equipped with an image slicer configured for 3 slices.

The observation dates of the star in the IRAS\,07430 system and the results of
radial velocity measurements are given in Table~2. The spectrum of one star from
the considered sample of C--rich stars, in the IRAS\,04296 system, and a part of
the IRAS\,23304 spectra, due to their weak apparent brightness, V=14.2$^m$ and
12.99$^m$, respectively [9], we have obtained with the PFES moderate-resolution
echelle spectrograph [42]  at the primary focus of the 6-m telescope. As follows
from the publications  [29]  and  [33], the spectra of these two stars are
saturated with absorptions of heavy metals of the s-process, the H$\alpha$ line has a
profile typical of post-AGB stars with the appearance of emission at certain times
of observations. The main feature of the spectrum of both stars, the emission in
the Swan bands of the C$_2$ molecule, is shown in Fig.~5 of Klochkova et al. [33]
in comparison with the powerful emission in the spectrum of the Egg~Nebula,
a bipolar protoplanetary nebula, associated with the post-AGB star V1610\,Cyg.
A weak emission in the Swan bands is also recorded in the spectrum of IRAS\,22223
(see Fig.~5 in the paper by Klochkova [21]. Note that  these stars with emission
in the Swan bands are surrounded by structured nebulae.

The study of variability in the spectra of cool stars (IRAS\,04296, IRAS\,07430 and
others with the apparent brightness of V$\ge 13^m$ ) is complicated by several factors.
Firstly, the visible brightness of such a star is the limiting brightness for
high-resolution spectroscopy. Therefore, when the seeing deteriorates, even over
four hours of signal accumulation, the $\rm S/N$, required, in particular, for
identifying numerous narrow rotational features of the Swan bands cannot be achieved
(see the (1;0) band in Fig.~1a).
Secondly, the high degree of blending due to the saturation of the spectrum with
strong ions absorptions, the equivalent widths of which are often comparable
with the intensity of H$\alpha$ or exceed it. The fragments of the spectrum presented
in Fig.~2 illustrate this well. A separate problem is the lack of reliable radio
spectroscopy data, which for the stars with envelopes allow one to reliably fix
the systemic velocity.
For example, observations by Hrivnak and Bieging [43], conducted for 12 post-AGB
stars in the (4--3) and (2--1) details of the CO molecule, common in envelopes,
did not yield any results  for IRAS\,07430.

\begin{table*}[ht!]
\medskip
\caption{Basic information on the sample of C-rich post-AGB stars associated with IR sources}
\begin{tabular}{l|c|c|c|c|c|c|c|c}
\hline
Parameter &\multicolumn{8}{c}{\hspace{0.5cm}IRAS sources:}  \\ 
\cline{2-9}
   & z02229&04296&07134&07430&19500&22223&23304 & References \\ 
\hline
$\pi$, mas& 0.3806 & 0.2366&0.4538&3.0577&0.3992 &0.3325&0.2366 &[27] \\ 
V, mag    &12.1    &14.2   &8.2   &12.8 & 8.7  & 9.7 & 13.1 &[9] \\ 
$\Delta$V, mag& 0.70&0.12  &0.24  &0.29 &0.16&0.26& 0.22 &[9] \\ 
Teff,\,K  & 5952  & 7272 & 7485  &5519&8239 &6008 & 6276 & [44] \\ 
  E(B-V)  & 1.90  &2.03 &0.43 &1.04 &0.56 &0.43&1.83 & [44]  \\  
\hline
\end{tabular}
\label{stars}
\end{table*}

\section{ANALYSIS OF THE MAIN RESULTS}

The major stage in the study of post-AGB stars has become the availability of  the Gaia
mission results and the Gaia DR3 catalog. Reliable parallaxes of stars, as a rule, provide
their distances, luminosities, initial and  current masses. Kamath et al.~[44], based
on the parallaxes of the Gaia~EDR3 catalog, conducted SED modeling for a sample of single
C-rich  stars at the post-AGB stage and obtained a set of fundamental parameters and
additional information for them. The fundamental parameters of the stars determined in this way,
in combination with the features of chemical composition
of the atmospheres and envelopes of evolved stars, provide a refinement of the
evolutionary stage and a reconstruction of the history of chemical composition
changes. In particular, the above-listed parameters were obtained by Kamath
et al.~[44] for all members of the sample of C-rich post-AGB stars we consider in this survey.

\begin{figure}[hbtp]
\includegraphics[angle=0,width=0.75\textwidth,bb=40 45 750 530,clip]{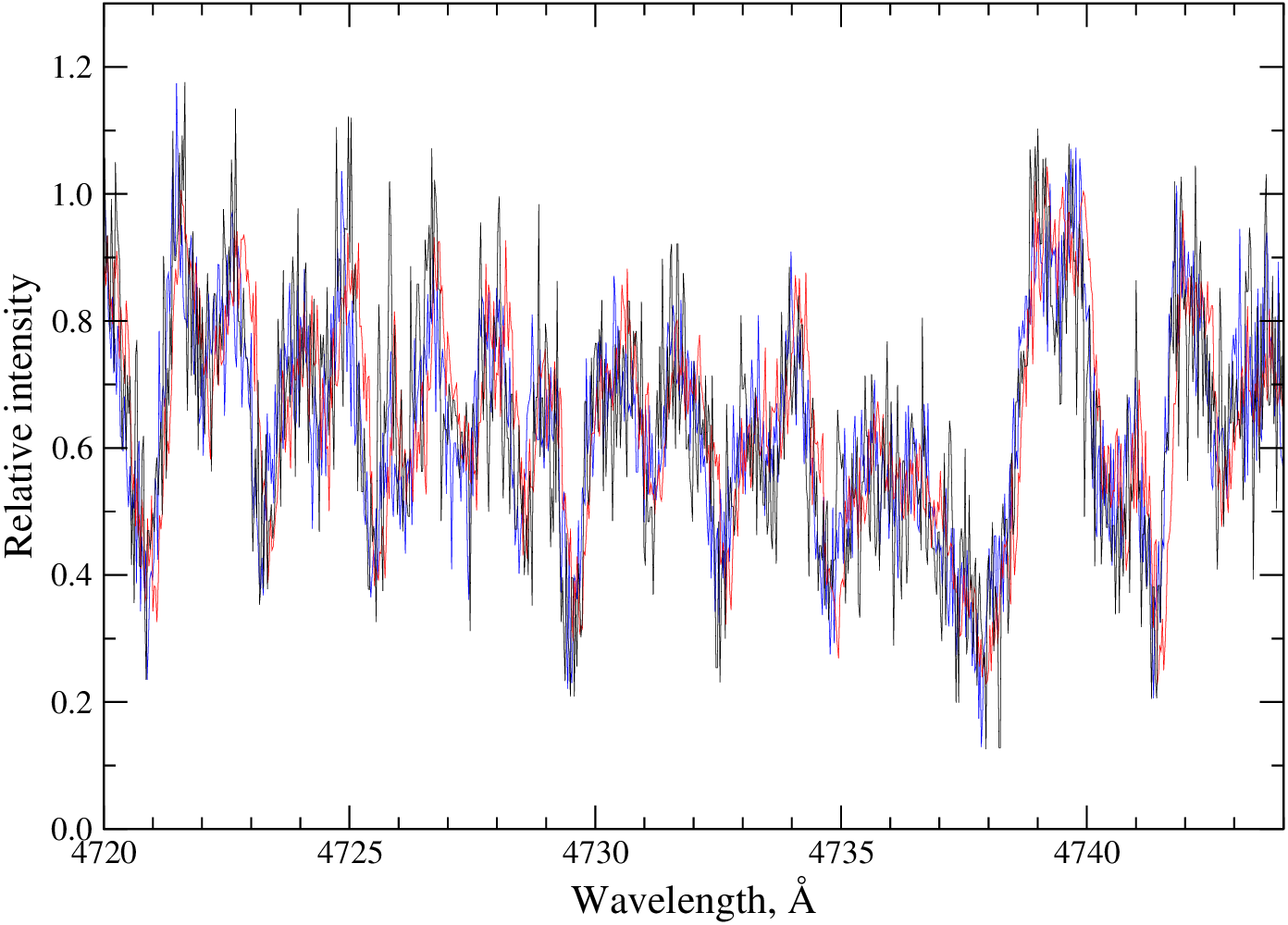}   
\includegraphics[angle=0,width=0.75\textwidth,bb=40 45 750 530,clip]{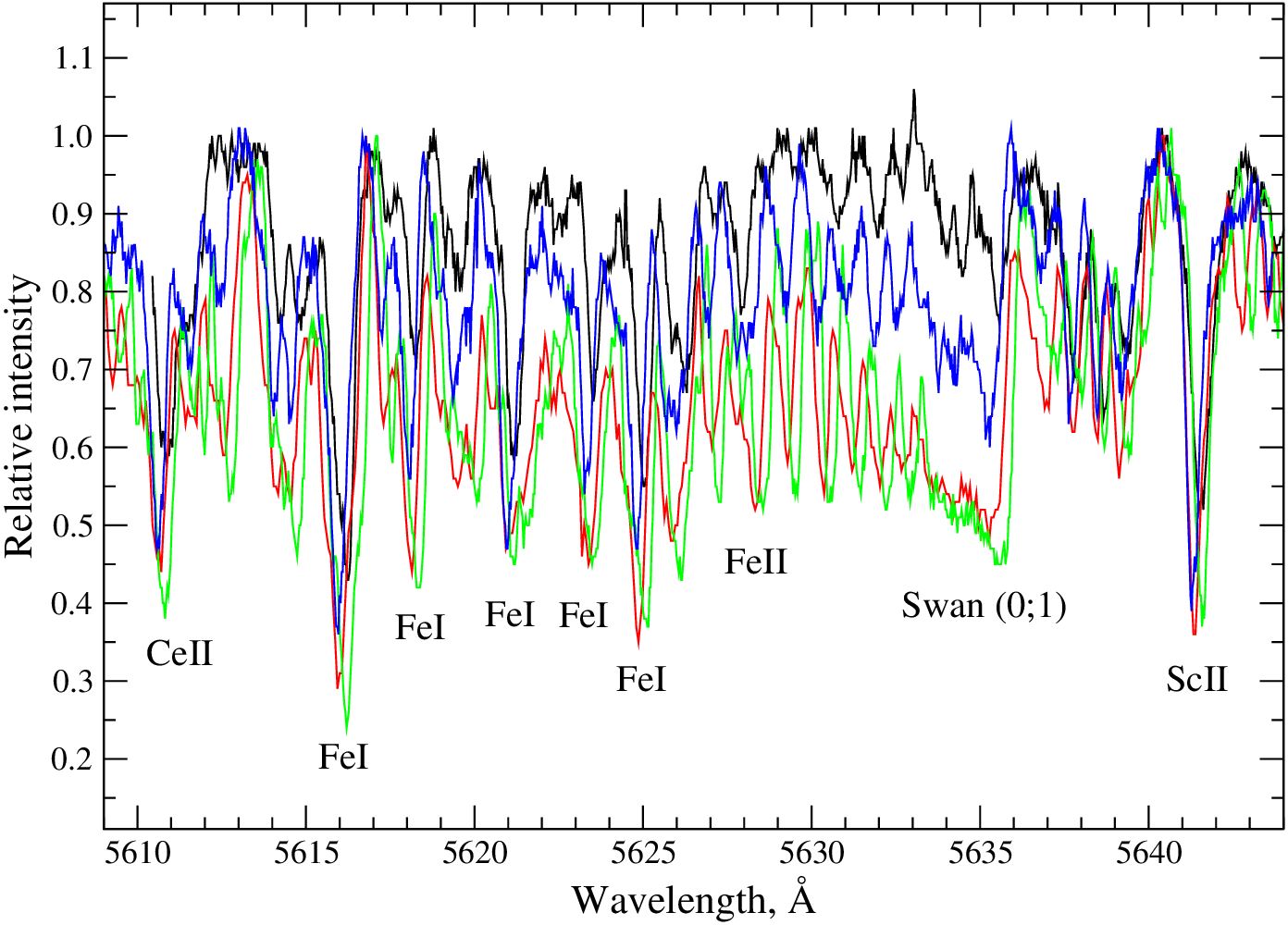}  
\caption{Panel (a) shows a fragment of the spectrum with the Swan band (1;0) in the spectra
of IRAS\,07430 obtained on different nights. Panel (b) gives a fragment of the spectrum with
the Swan band (0;1) in the spectra of IRAS\,z02229 obtained on different nights.
 The identifications of a number of absorptions are indicated}
\label{Swan}
\end{figure}

\begin{figure}[hbtp]
\includegraphics[angle=0,width=0.7\textwidth,bb=40 40 710 530,clip]{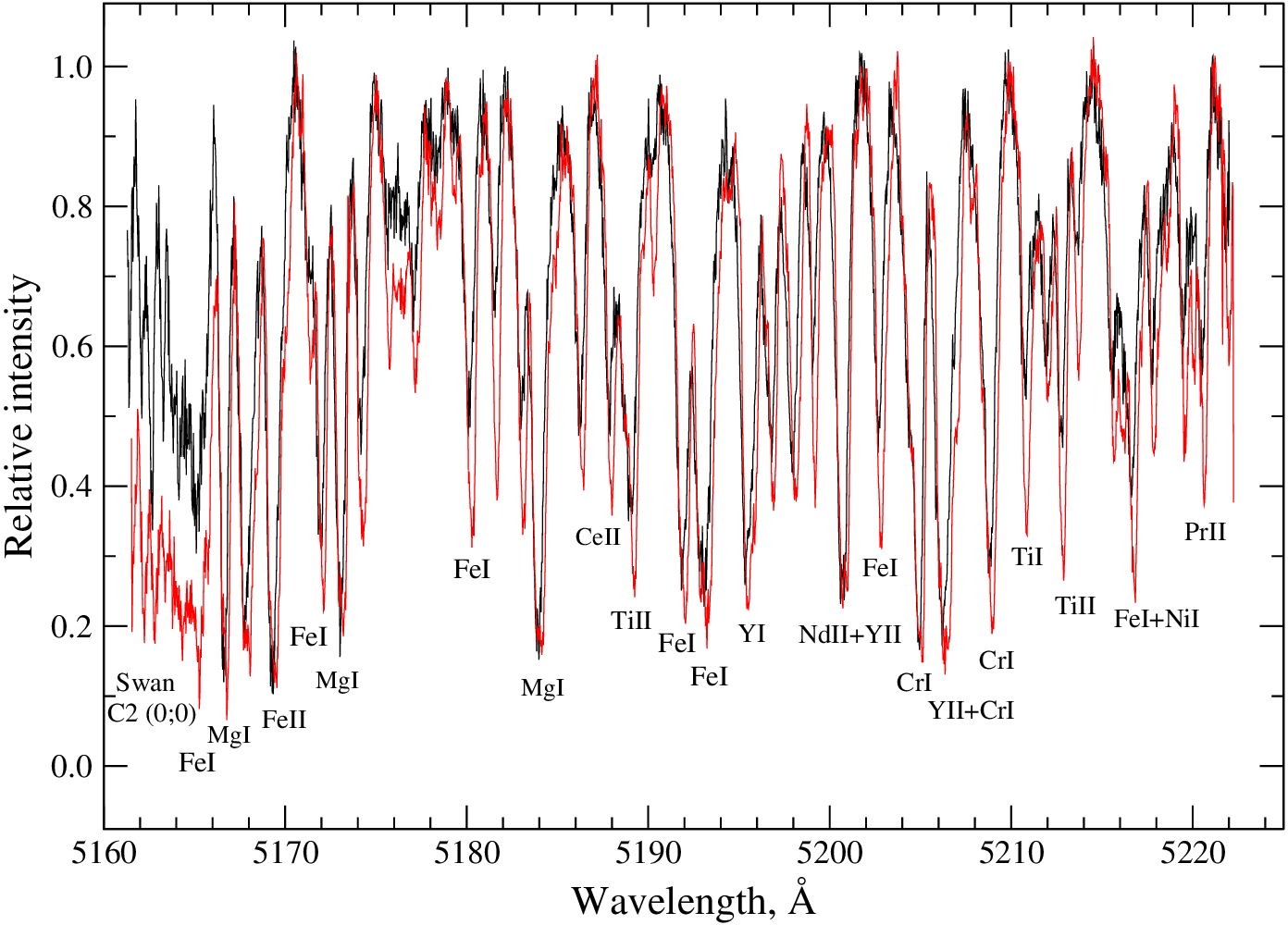} 
\includegraphics[angle=0,width=0.7\textwidth,bb=40 40 710 530,clip]{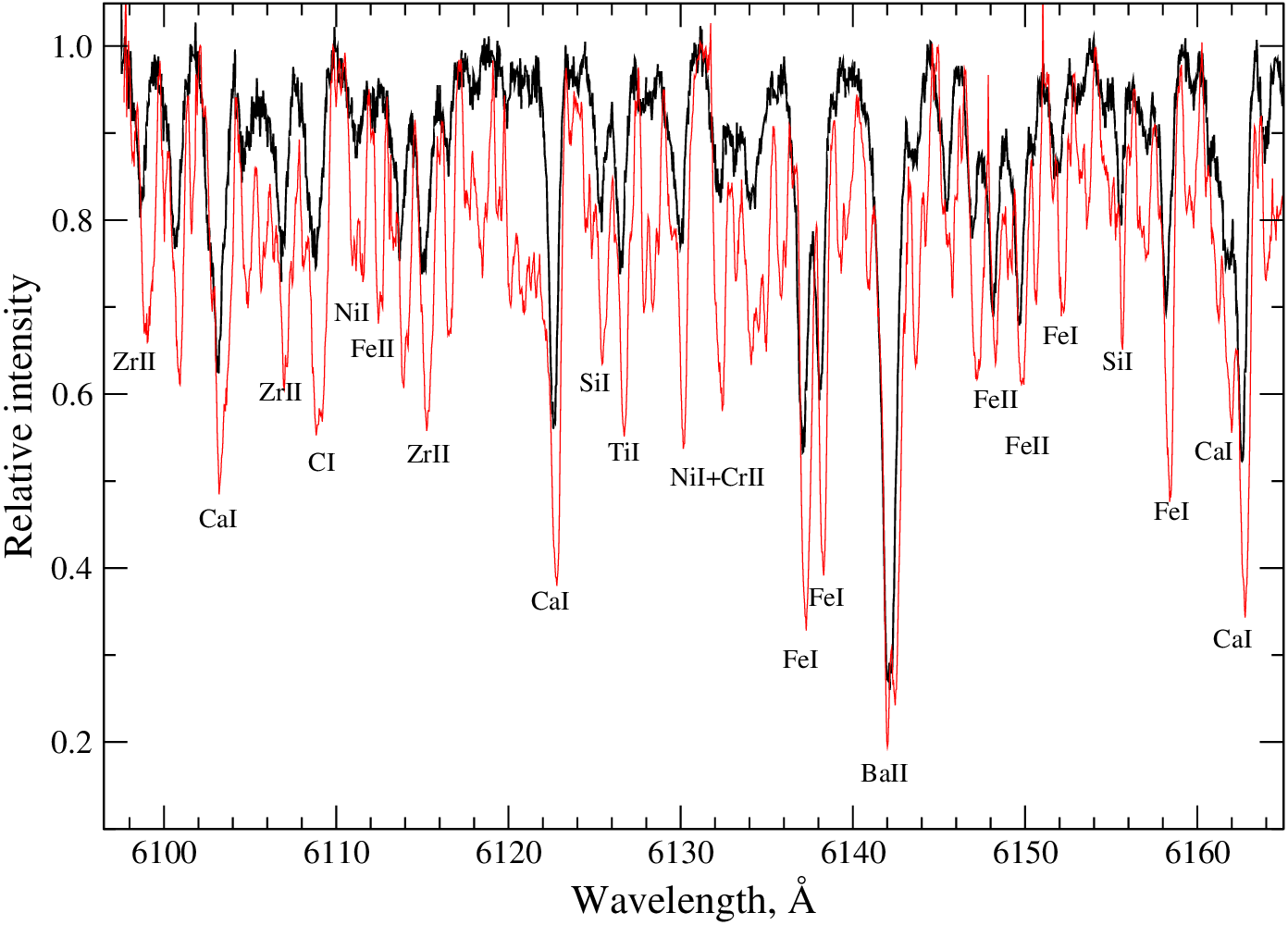} 
\caption{The fragments of the spectra of IRAS\,07430 (black line) and IRAS\,z02229 (red).
The observed wavelengths are shown along the abscissa. The identification of the main
absorptions is marked. In the lower panel, we see the splitting of the Ba\,II$\lambda$6141\,\AA{}
absorption core in the spectrum of IRAS\,z02229, which is absent in the spectrum of IRAS\,07430.}
\label{Fragm}
\end{figure}

{\bf Luminosity of the central star of IRAS\,07430.}
 In the work of Kamath et al. [44] we are particularly interested in the
result concerning the status of the central star of the IRAS\,07430 source, for which
the authors of the paper determined an extremely low luminosity ${\rm L/L_{\odot}}$=20.
At the same time, the luminosity of the kindred star in the IRAS\,z02229 system is more
than two orders of magnitude higher. ${\rm L/L_{\odot}}=12959$~[44].
It is curious that with such a difference in luminosity, both stars have almost the same
equivalent width  of the oxygen triplet O\,I $\lambda$7774, which is considered to be a
good criterion for the absolute magnitude of G-supergiants. In the spectra of IRAS\,z02229
and IRAS\,07430, W$_{\lambda}$(OI\,7774)=0.99 and 0.91\,\AA{}, respectively. Using
the M$_v$--W$_{\lambda}$(OI\,7774) calibration dependence Kovtyukh et al~[45],
we  obtain for IRAS\,07430 the absolute magnitude of M$_v\approx-3.2^m$  and the logarithm
of the luminosity log L$/L{\odot}\approx$3.19 within the range of values for post-AGB
stars. For example, according to~[44], the related star of the IRAS\,22223  source has
a close luminosity: log L$/L{\odot}\approx$=3.34.

We consider the obtained luminosity estimate for IRAS\,07430, typical for post-AGB
stars, to be an important result in the analysis of the set of systemic parameters
due to the presence, according to Kamath et al.~[46], of a strong relationship
between the luminosity of post-AGB stars and their mass and the initial mass of their
progenitors on the MS. Kamath et al.~[46], modeling in detail the stages of
evolution  at the AGB and post-AGB stages, have established a lower mass threshold:
stars undergoing the third mixing originate from the progenitors with masses in the
range of 0.85$\div0.95\mathcal{M}_{\odot}$ at the beginning of the AGB. Judging
by the chemical composition of  the atmosphere, the central star in the IRAS\,07430
system has undergone  the third mixing: the chemical composition of its atmosphere is
similar to that of IRAS\,z02229, given similar parameters of the model atmosphere.
Therefore, we have a paradoxically low luminosity of IRAS\,07430, estimated by
Kamath et al. [44]  using the parallax of Brown et al.[27]. Kamath et al. [44]
also point at the low quality of astrometric data for IRAS\,07430, for which the
value of the parameter RUWE=21.8 is the highest among the stars in the sample,
while the good quality data have RUWE$\le$1.4. We also note the impossibility of
estimating the luminosity of the star in the IRAS\,19500 system, since in the
spectrum of this fairly hot star (Teff=8239$\pm$250\,K according to [44]  the
intensity of the triplet (its W$_{\lambda}$(OI\,7774)=2.1\,\AA{} is beyond the
calibration limits in~[45].

\begin{table*}[ht]
\medskip
\caption{The results of measurements of the heliocentric radial velocity Vr in the spectra
in the IRAS\,07430 system. The number of spectral details measured to determine the average
Vr value for each observation date is indicated in parentheses }
\begin{tabular}{ c| l|  c|  c  }
\hline
Date & \multicolumn{3}{c}{\small Vr, km/s} \\  
\cline{2-4}
        &sorptions &   H$\alpha$(core) &Swan    \\  
\hline
   1    & \hspace{7mm}  2 \hspace{7mm} &   3 & \hspace{7mm} 4   \\  
\hline
19-23.12.1996$^1$ & 35$\pm 1$ & & 22$\pm 1$\,(23),  24$\pm 1$\,(12) \\ 
\hline
06.04.2018 & 36.6$\pm0.14$(268)&40.5   &21.4$\pm 0.6$\,(11)   \\ 
\hline
11.04.2018 &36.2$\pm 0.12$\,(380)&  41.2 & 20.2$\pm 0.3$\,(28) \\ 
\hline
07.12.2019 &37.7$\pm 0.16$\,(488)&  43.8 &20.0$\pm 0.2$\,(28) \\ 
\hline
27.03.2024 &39.2$\pm 0.14$\,(194)& 42.3 & $ 20.7\pm 0.2$\,(55) \\ 
\hline
\multicolumn{4}{l}{\footnotesize $^1$ -- means Vr for 1996 based on the data in ~[36]} \\[-10pt]
\end{tabular}
\label{velocity}
\end{table*}

\subsection{Features of the Spectra and Radial Velocity Patterns}

All the kinematic features (pulsation amplitude A$_{\rm Vr}$, splitting (or asymmetry) of strong metals
absorptions, presence  of stratification of velocities in the stellar atmosphere) revealed in the
atmospheres of the considered C-rich post-AGB stars are collected in Table~3.
In this subsection we will focus on some of the features of individual stars studied.

{\bf IRAS\,z02229 and its closest analogues.} Let us start with the features  of the spectrum and
velocity field of one of the less studied objects of the considered sample of C-rich stars, namely,
IRAS\,z02229. Klochkova and Panchuk [38]  found  several expected features in the spectra of
this typical C-rich star: the variability  of the H$\alpha$ profile shown in Fig.\,3 and the
splitting of the absorption profiles of metals and ions (K\,I, NaI, YII, ZrII, BaII, LaII, CeII, NdII)
with a low lower-level  excitation potential ($\chi_{low}\le 1$eV). Their profiles have asymmetric
shapes or are split into two components with different velocities. An important point is that in the
presence of an obvious splitting of absorptions, the position of their long-wave component  coincides
with the position in the spectrum of unsplit absorptions of other metals, which  confirms its formation
in the stellar atmosphere. The position of the short-wave component is close to the position of the
features of the Swan system bands, which indicates its  formation in the circumstellar envelope.
Note that Fig.\,1 of  Za\v{c}s and Pu\c{k}\={\i}tis [47]  shows a fragment of a high resolution
spectrum in  the range of 6120--6150\,\AA{}, containing the split absorption of BaII\,$\lambda$6141,
which is noted by the authors  of this paper.

\begin{figure}[hbtp]
\includegraphics[angle=0,width=0.4\textwidth,bb=10 80 550 680,clip]{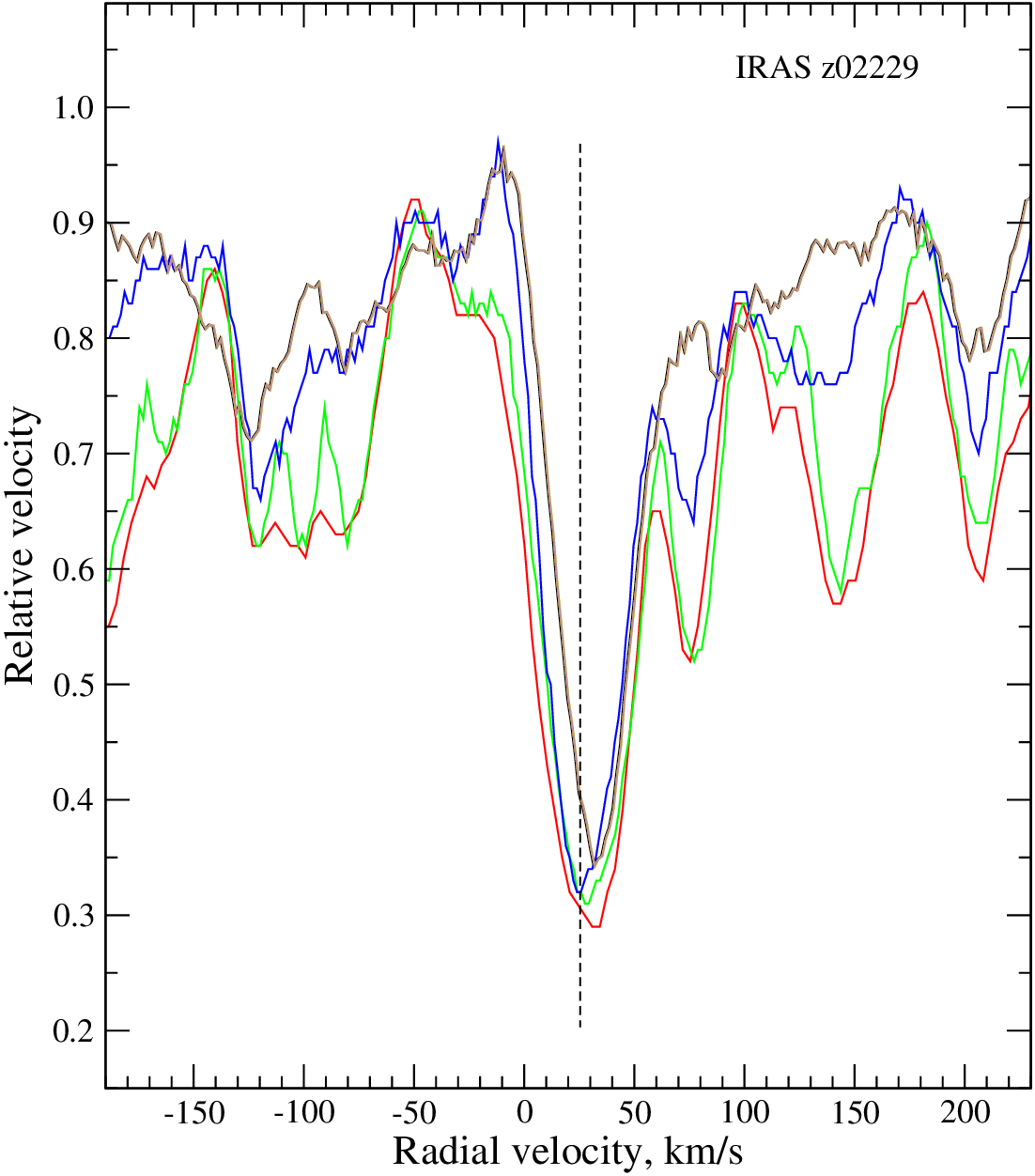}  
\includegraphics[angle=0,width=0.4\textwidth,bb=10 80 550 680,clip]{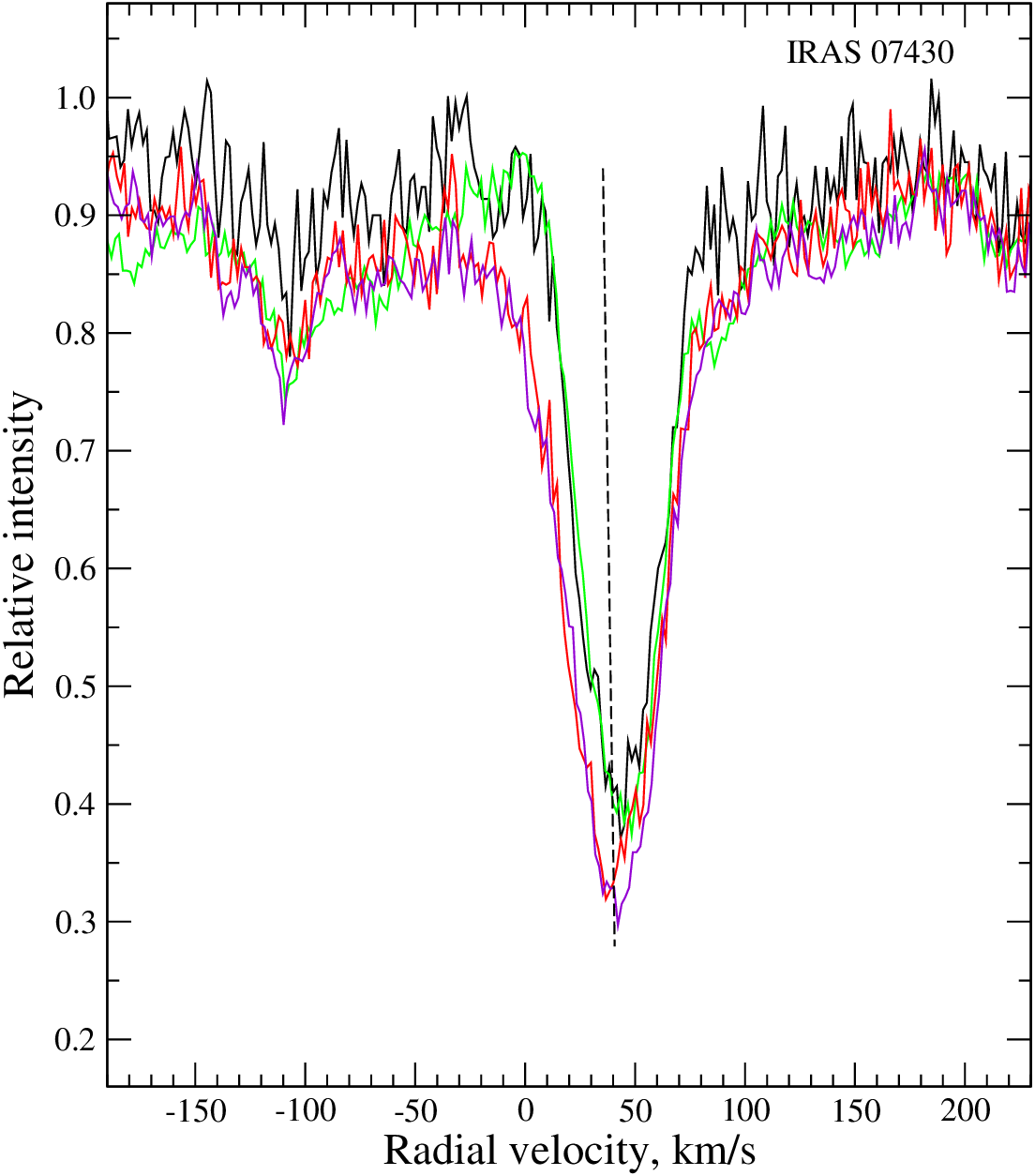}  
\caption{The H$\alpha$ profile in the ``radial velocity--relative intensity'' coordinates in the spectra
    obtained on different dates. The position of the dashed vertical corresponds to
   the value of the systemic velocity Vsys=+24.3\,km/s for IRAS\,z02229  and the value
   Vsys=+37.0\,km/s for IRAS\,07430 according to the radio spectroscopy data of Hrivnak and Kwok [48].}
\label{Halpha}
\end{figure}

 The splitting of strong absorptions in the spectra of V5112\,Sgr looks even more complex.
 As it is clearly seen in Fig.~4, their profiles are split not
into two, but rather into three components. In addition, a peculiar variability of these
profiles over time is observed (see Fig.\,3 in the article~[31).
As follows  from a comparison of the data for the available observation dates, the positions
of long-wave components of the split absorptions change synchronously with the positions
of symmetric absorptions in the spectra of the star. This means that the long-wave component
of the BaII\,lines is formed in the atmosphere of the star. The width of both short-wave
components formed in the two-layer envelope is significantly smaller than that of the
atmospheric ones, and their positions are  stationary (see Fig.\,3 in  [31]), which confirms
their formation in the circumstellar medium.

Previously, a similar splitting  (or asymmetry of the profile in the form of an elongated
shortwave wing) of strong absorptions  was found in the spectra of several post-AGB stars
with envelopes. This information is presented in Table~3. In the spectra of three post-AGB
stars with similar parameters in the IRAS\,07134 (Klochkova,
[30] and  [28]), IRAS\,22223~[32]  and IRAS\,22272+5435 (hereafter IRAS\,22272)
(Klochkova et al., [49] 2009) systems there is no obvious splitting of absorptions in
the spectra, but an asymmetry of the profiles of strong absorptions is registered.

\begin{figure}[hbtp]
\includegraphics[angle=0,width=0.4\textwidth,bb=20 20 560 680,clip]{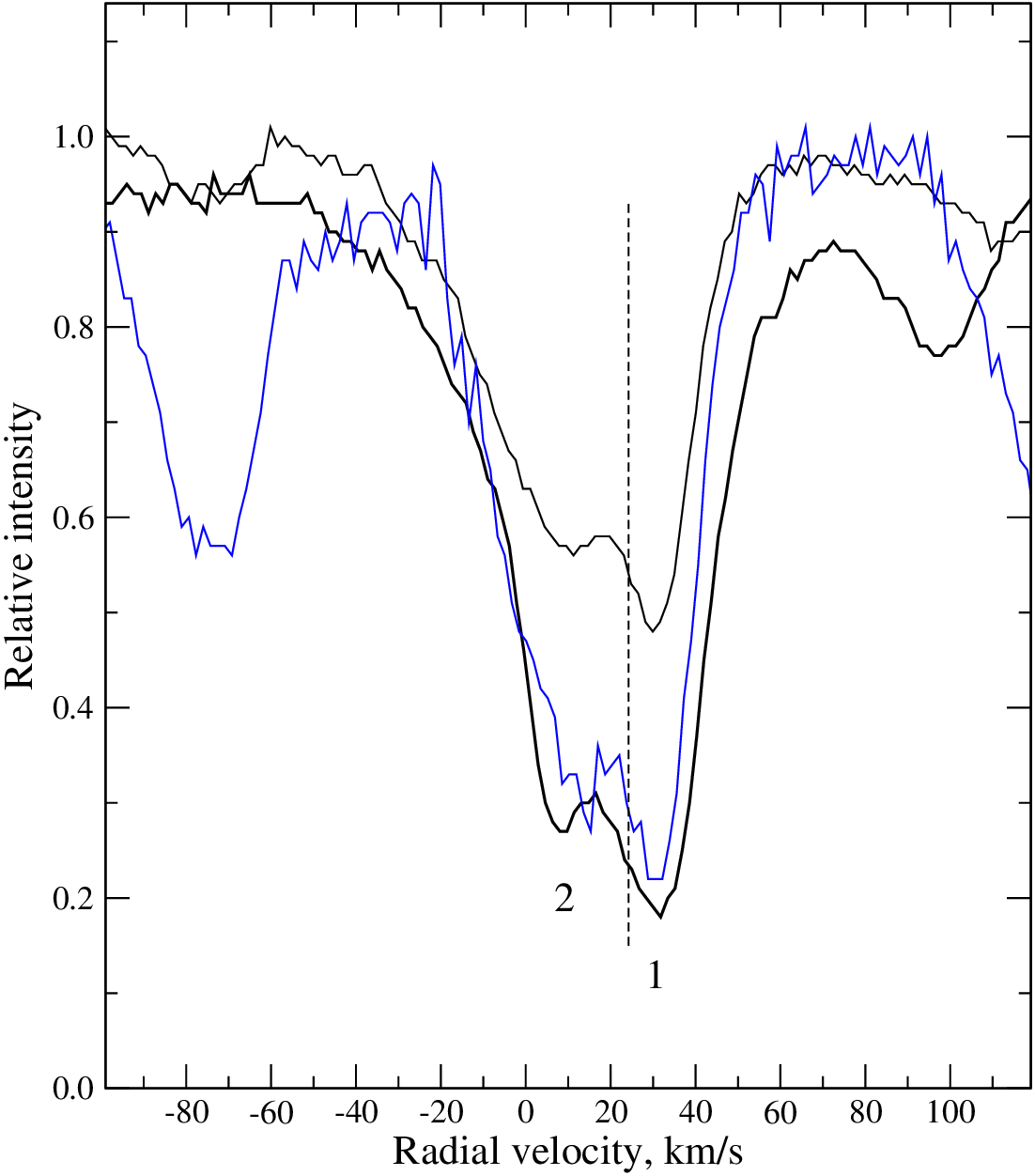} 
\includegraphics[angle=0,width=0.4\textwidth,bb=20 20 560 680,clip]{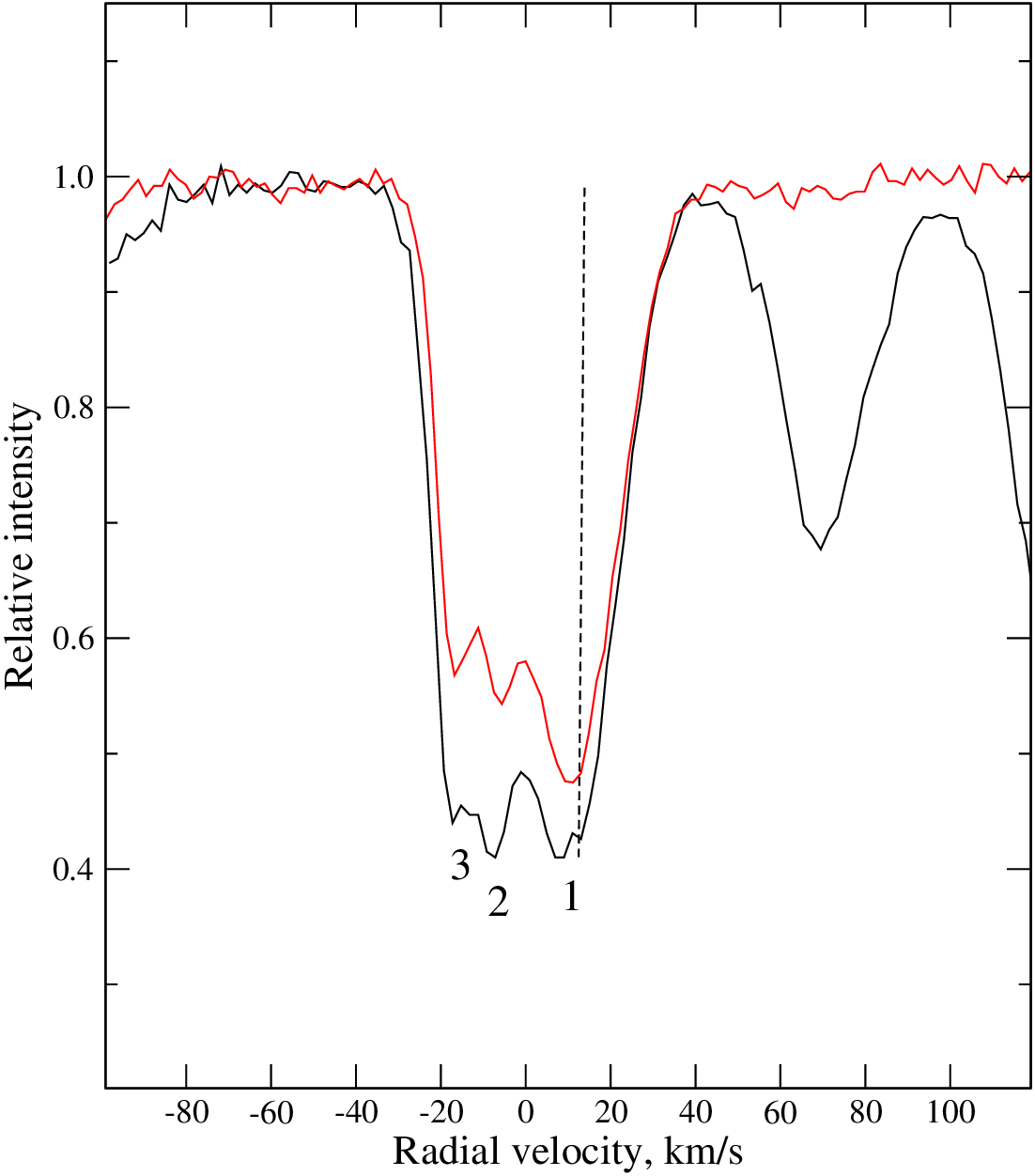} 
\caption{The split profiles of heavy metal absorptions. The left plot gives the spectrum of
IRAS\,z02229: BaII\,6141\AA{} drawn by the thick black line, Y\,II\,5200\,\AA{}  --  the blue
line and LaII\,6390\,\AA{} -- the thin black line. Component ``1'' is the atmospheric absorption,
the shortwave component ``2'' is from the envelope. The right plot presents the spectrum of
IRAS\,19500: BaII\,4554\,\AA{} --  the black line, BaII\,4934\,\AA{} -- red. The position of the
     dashed vertical line indicates the value of the systemic velocity.}
\label{split}
\end{figure}

{\bf V5112\,Sgr}. Of particular interest in studying the  features of the kinematic state
of the atmosphere is the post-AGB star V5112\,Sgr in the system of the IR source IRAS\,19500,
which is also included in the list of single C-rich-stars in the paper of Kamath et
al.~[44]. In the paper by Klochkova [31], based on six spectra obtained on the
BTA\,+\,NES on different dates over 1996--2012, radial velocity variability was found based on
symmetric absorptions of metals with no signs of anomalies: the average value
Vr(abs)=12.5\,km/s with a standard deviation of 2.5\,km/s. The spectrum of this star
shows a significant variability  of the H$\alpha$ profile. Fig~1 in the paper by
Klochkova~[31]  illustrates the variations in the H$\alpha$ profile from the inverse
P\,Cyg-type to a profile with two emission peaks. This variation occurred in parallel
with a significant variation in the photometric characteristics of the star, registered
by Hrivnak et al.[50]  during the long-term monitoring.

{\bf  IRAS\,04296}. Of interest is also the IR source system IRAS\,04296, which, like
IRAS\,z02229, has a powerful excess of IR flux. The central star in this system has
advanced further to the planetary nebula stage, but in general its fundamental parameters,
details of the chemical composition of the atmosphere, and  SED in the of this system
(Klochkova et al.~[29]) are close to those of the IRAS\,z02229 system. The star
in the IRAS\,04296 system is weak for the high-resolution spectroscopy
(V$>14^m$, B$>16^m$), but its optical spectrum (spectral class G8\,Ia) is similar to
the spectra of other stars in our sample. In the IR range, the spectrum contains all
the features inherent in C-rich post-AGB stars, including the known, but not yet fully
identified emission at 21\,$\mu$~[50]. The object is a record holder  for the emission
intensity in the Swan system bands (0;0) and (0;1) of the C$_2$ molecule ([26], [51]).
It can be assumed that the greater distance of this object from the AGB phase probably
leads to a significant expansion of the envelope, which contributes to the formation
of emission in the circumstellar features.

{\bf IRAS\,07430}. In recent years, it became possible to obtain and analyze the spectra
of the weak central star of the IRAS\,07430 source. The main features of the optical spectrum
of IRAS\,07430 are illustrated in Figs.\,1--3. In each spectrum of IRAS\,07430 that we
have obtained, the radial velocities Vr were measured from the positions of photospheric
absorptions, the rotational features of the Swan bands, and the circumstellar NaI and K\,I
features. The results of the measurements are presented in Table~2.
From date to date, the velocity based on the photospheric absorptions varies around the
average value Vr=37.0\,km s−1 with a standard deviation of $\Delta$Vr$\approx$0.8\,km/s,
which may be a manifestation of low-amplitude pulsations in the stellar atmosphere or the
presence of spots on its surface. The position of the H$\alpha$ core varies in a small
interval of 40.5$\div$43.8\,km\,s. The variability of the profile of this line may be
caused by a weak wind, due to which its short-wave wing changes (see Fig.\,3b). However,
this conclusion requires new observations with a high S/N ratio.

Based on the positions of the rotational components of the Swan bands (0;0), (1;0), and (2;0)
of the C$_2$ molecule formed in the circumstellar envelope, the average radial velocity
of about Vr(Swan)$\approx$21\,km/s was determined and the envelope expansion velocity
Vexp$\approx$16\,km/s was estimated, typical for stars of this type. The results of measuring
Vr based on the positions of the rotational features of the Swan bands (see Table 2) lead to
the conclusion about the absence of velocity variability in the stellar envelope. The stability
of the Swan band (1;0) in the spectra of IRAS\,07430 is also illustrated by the upper panel
of Fig.\,1a. At that, the Swan band (0;1) in the spectra of IRAS\,z02229, shown in panel (b)
of the same figure, has a clearly variable intensity.  Somewhat later, the authors~[52]
continued their study of the variability of the spectrum of IRAS\,z02229
using high-resolution spectra. In the long-wavelength region of the spectrum, they have also
recorded a variability of the features of the CN and C$_2$ molecule bands, formed in the
circumstellar medium and confirmed the pulsations of the star with a period of about 154 days.
For a star in the IRAS\,z02229 system, the half-amplitude of the velocity variability based
on the photospheric absorptions is $\Delta$Vr$\approx 1.4$\,km/s~[38], which is higher than
this parameter $\Delta$Vr$\approx 0.8$\,km/s in the IRAS\,07430 system.
Therefore,  based on the available spectroscopic monitoring data a conclusion was made about
the difference in the kinematic state of the atmospheres of these two kindred C-rich stars.
Additional confirmation of the stability of the atmosphere of IRAS\,07430 is the absence of
radial velocity stratification in it. We have searched but failed to find in the spectra of
IRAS\,07430 any dependence of the radial velocity values Vr(abs) on the absorption intensity,
as found for IRAS\,z02229 in [38]. For illustration, Fig.\,5 presents the velocity measurement
data from the absorptions of different depths in the spectrum of the star for March~27,~2024.
The absence of a significant stratification of radial velocity in the atmosphere is indicated
by the independence of Vr(abs) from the level of line formation in the atmosphere.

\begin{figure}[ht]
\includegraphics[angle=0,width=0.55\textwidth,bb=20 35 710 570,clip]{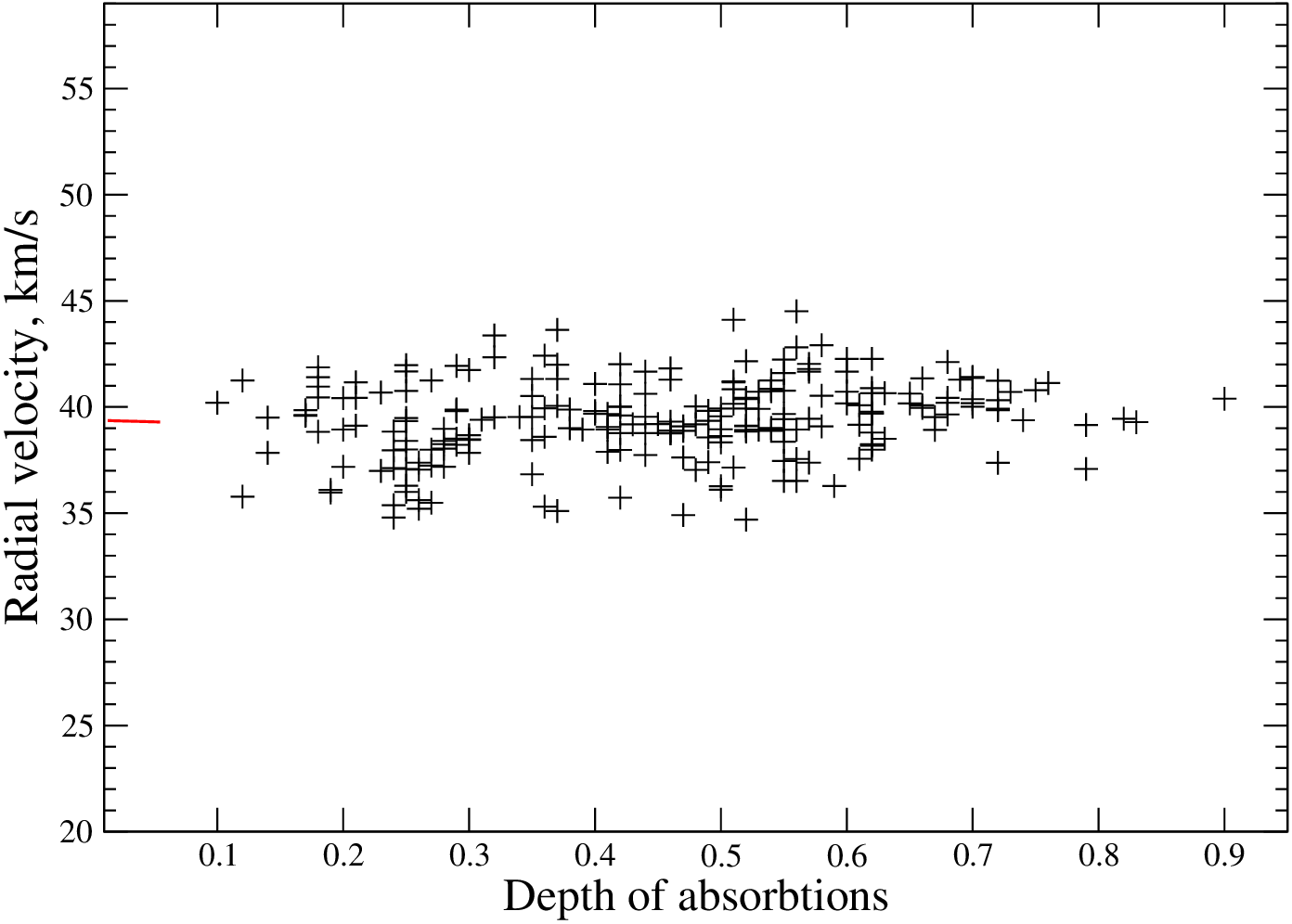}  
\caption{Radial velocity values, Vr(abs), measured from the absorptions in the spectrum of IRAS\,07430
obtained on May~27~2024.  The short red line indicates the averaged velocity Vr(abs)=39.2\,km/s
for this date based on the set of absorptions. }
\label{Vr_depth}
\end{figure}

{\bf IRAS\,22223 and IRAS\,22272}. Previously, radial velocity variability with an amplitude
of   $\Delta$Vr$\approx$1--2\,km/s was detected in the BTA\,+\,NES spectra for the post-AGB
star V448\,Lac in the IRAS\,22223 system by measuring the positions of weak absorptions~[32]).
This star is a full member  of the subgroup of post-AGB stars  with detached envelopes,
double-humped SED, and atmospheres enriched in carbon and heavy metals. Klochkova et al.~[32]
recorded numerous manifestations of instability in the spectra of this star: variability of the
H$\alpha$ profile, asymmetry and variability of the profiles of strong absorptions with a low
excitation potential (primarily these are variable profiles
of Ba\,II absorptions), as well as emission in the Swan band  (0;1)\,5635\,\AA{}. The variability
of the Ba\,II profiles is due to the appearance of an emission component formed in the envelope,
and at the same times, emission is also recorded in the Swan band (0;1)\,5635\,\AA{} of the
C$_2$ molecule.  In addition, differential shifts in the Vr pattern were detected --  the difference
in radial velocities from absorptions of different depths reaches the values from 0 to 8\,km/s
 on different observation nights (see Fig.\,8 in the paper~[32]).

As follows from the publication by Kamath et al.~[44], the group of single C-rich post-AGB
stars also includes the source IRAS\,22272. The main parameters of HD\,235858, the central star
of this system, are close to the average values for the sample. The apparent magnitude of 9.5$^m$,
an insignificant distance and low absorption make this star accessible for high-resolution spectroscopy.
The results of  the features of the radial velocity pattern in its atmosphere and
circumstellar envelope~[47] were published. Based on the long-wavelength spectra of the source
IRAS\,22272, authors~[47] detected variability in the bands of carbon-containing molecules CN and C$_2$.
At certain phases of observations, emissions with an intensity of about 10\% above the continuum and
with a variable position appear in the bands of these molecules. The combination of these features
can be explained by the presence of a non-spherical circumstellar envelope formed owing to the
summation of stellar matter flows due to winds blowing at different velocities at the AGB and post-AGB
stages (Ueta et al.~[53]).

Previously, emission in the Swan bands of the C$_2$ molecule was recorded in
the optical spectra of IR sources: IRAS\,04296~[29], IRAS\,08005$-$2356~[54],
IRAS\,22223~[32], IRAS\,23304~[35]. Images of these IR sources obtained using the HST
space  telescope have asymmetric and, as a rule, structured shells (Ueta et al., [55] 2000).
The maximum efficiency of emission in the Swan bands was registered by Klochkova et al.~[56]
in the spectrum of the nebula RAFGL\,2688.

\begin{figure}[hbtp]
\includegraphics[angle=0,width=0.5\textwidth,bb=20 70 560 680,clip]{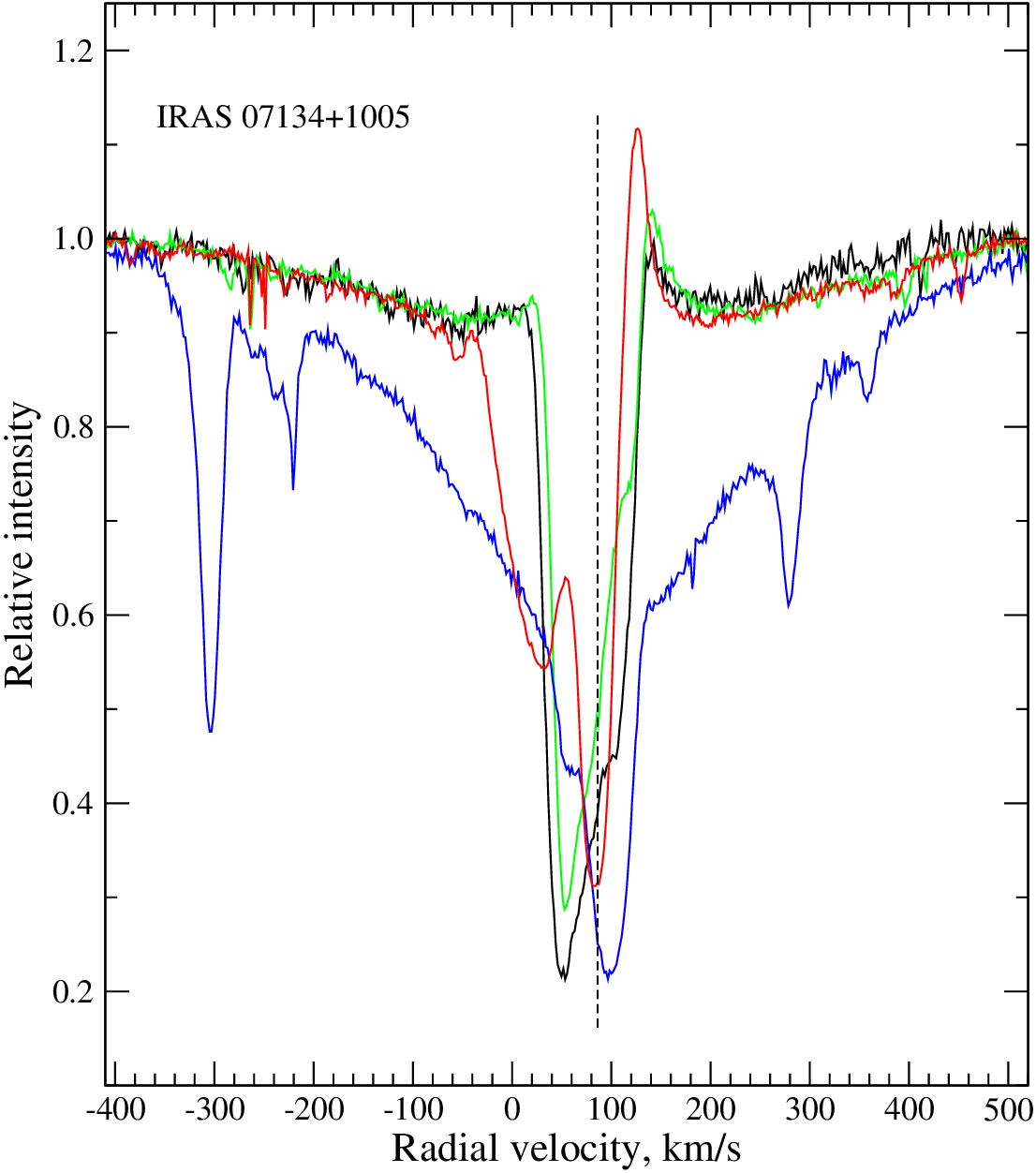}  
\caption{The H$\alpha$ profiles in the spectra obtained at the BTA with the NES spectrograph: May~5,~2007 --
the black line, October~11,~2013 -- green, March~30,~2024 -- bred. The H$\beta$ profile in the spectrum
for March~30,~2024 is shown in blue. The position of the dashed vertical corresponds to the systemic
velocity of Vsys=86\,км/с by  Bujarrabal et al~[57].}
\label{IRAS07134_Ha_Hb}
\end{figure}

{\bf HD\,56126 -- a canonical object at post-AGB stage}. The kinematic features of a fairly
bright star HD\,56126 in the IRAS\,07134 system were studied in detail using a large volume
of high-resolution spectra by Barthe`s et al.~[58] (2000), who found pulsations with a
half-amplitude of 2.7\,km/s. The monitoring of this star was later continued by Klochkova and
Chentsov [59], who, using the spectroscopy data from the BTA from 1993 to 2005,
presented a variety of H$\alpha$ profiles of the P\,Cyg type, changing from a direct
P\,Cyg to an inverse one (see Fig.\,3 in the article~[59]).
The half-amplitude of velocity variations based on weak absorptions (with a residual intensity
close to the local continuum level) is 2--3\,km/s, confirmed by the stability of the
expansion velocity of the circumstellar envelope of HD\,56126, recorded from the C$_2$
and NaI lines. In addition, a different behavior of velocity was found based on the lines
of different degrees of excitation, which are formed at different depths in the stellar
atmosphere. Figure~7 in the paper~[59] demonstrates both effects: a manifestation of
pulsations and a stratification of radial velocity, based on the absorptions of different
intensities.

The variability of the complex H$\alpha$ profile in the spectrum of the fairly bright post-AGB
star HD\,56126, identified with the IR source IRAS\,07134, has been studied by many authors.
A large collection of H$\alpha$ profiles in the spectrum of IRAS\,07134 based on eight years of
observations was published by Barth\`{e}s et al.~[58] and Lebre et al.~[60], who concluded
that there is a significant and complex variability in the absorption part of the H$\alpha$ profile,
caused by the passage of waves in the atmosphere of the star.

In recent years, several new spectra of IRAS\,07134 have been obtained with the NES spectrograph; in
general, the set of features is consistent with previous observations, presented in detail in the
works~[59]. At the same time, a significant variation in the H$\alpha$
profile in the spectra of three dates over the last years of our observations is noteworthy, which is
clearly visible in Fig.\,6. This figure shows the systemic velocity Vsys=86\,km/s, determined for
IRAS\,07134 from radio spectroscopy data in the bands of the CO molecule~[57].
Note that for the first time over several decades of observations of this star, such an unusual  for
this star H$\alpha$  profile was recorded only in the spectrum of 2024. The profile contains, along
with the usual for post-AGB stars broad absorption wings, two emission ``shoulders'' and an absorption
core split into two components. The position of the long-wave component is consistent with the systemic
velocity, and the short-wave component is shifted by 60\,km/s. Barth\`{e}s  et al.~[60]  noted that
the presence of a structure in the H$\alpha$ absorption core can be caused by the presence of layers
with different velocities, since the coexistence of several shock waves in the region of formation of
hydrogen lines is not excluded. Note that the H$\beta$ profile in the same spectrum does not contain
significant peculiarities.

\begin{table*}[ht]
\medskip
\caption{Luminosity and kinematic features of the atmospheres of C-rich post-AGB stars
     associated with IR sources.}
\begin{tabular}{l|r|c|r|r|r|r|c}
\hline
Parameters &\multicolumn{7}{c}{\hspace{0.5cm}IRAS sources:}  \\ 
\cline{2-8}
                       &z02229&04296  &07134   &07430    &19500   &22223 &23304   \\ 
\hline
L/L$_\odot$~[44]      &12959  &10009  &5505     & 20     & 2163    &7054     &7712 \\
L/L$_\odot$\,(OI)$^1$ &1853   & ---   &11690    &1541    & ---     &9720     &8090 \\
A$_{\rm Vr}$, km/s    &1.4 [38]& ---  &2.7  [58]&0.8$^1$ &2.5 [31] &1$\div$2~[32]& ---  \\
Splitting/asym        &yes [38]& ---  &asym [59]&no$^1$  &yes [31] &asym [32]&yes [35] \\
Stratification        & no [38]& ---  &yes  [59]&no$^1$  &---      & yes [32]& --- \\
\hline
\multicolumn{8}{l}{\footnotesize{$^1$ --  Author's data. A dash means that there is no information}} \\[-10pt]
\end{tabular}
\label{kinematic}
\end{table*}

In conclusion, let us note that the group of single C-rich stars also includes another star,
rather weak (V=13.4$^m$) in the visible range for the spectroscopy on the BTA+NES, a star
in the IRAS\,20000+3239 system. Klochkova and Kipper [61], having only one spectrum of
this star obtained with the PFES spectrograph, revealed the presence of C$_2$ and CN
molecular bands, determined the model parameters, calculated the low metallicity
[Fe/H]$_{\odot}$=$-1.4$ and a detailed chemical composition of the atmosphere, typical of
C-rich post-AGB stars. Until now, repeated observations have not been carried out, and
unfortunately, there is no information on the velocity field variability. This task
requires high-resolution spectroscopic monitoring of stars in the IRAS\,04296
and IRAS\,20000+3239 systems.

Let us also briefly touch upon an issue that is important for the entire sample of C-rich post-AGB
stars, related to the identification of the absorption near $\lambda$6707\,\AA{} in their spectra,
which was often attributed to lithium, LiI. Due to the importance of the problem of lithium origin
in the picture of the chemical evolution of the Galaxy, researchers always pay special attention to
searching for this feature in the spectra of the stars they examine. For example, Reddy et al.~[36], 
Klochkova et al.~[62]  identified this absorption in the spectra of post-AGB stars with the LiI\,6707.76\,\AA{}
line and concluded that there is an  excess of lithium in the atmospheres of these stars. 
However, Reyniers et al.~[63] showed that in the spectra of C-rich post-AGB stars,
the absorption at 6707.8\,\AA{} should be identified with the absorption of CeII\,6708.099\,\AA{}.
Reddy et al. [36] (1999) also noted the unreliability of determining the lithium abundance in  the
atmospheres of both studied stars based on the intensity of the absorption near $\lambda$6707 due to
possible blending by the CN and CeII lines.  In our spectra of the studied C-rich post-AGB-stars,
we reliably identified this absorption with the CeII\,6708.099\,\AA{} line, since in the high-quality
spectra of these stars, radial velocity measurements based on the position of the controversial feature
6707\,\AA{}  are in excellent agreement with the average value of Vr based on numerous metal absorptions.

\begin{figure}[hbtp]
\hbox{
\includegraphics[angle=0,width=0.5\textwidth,bb=30 20 710 530,clip]{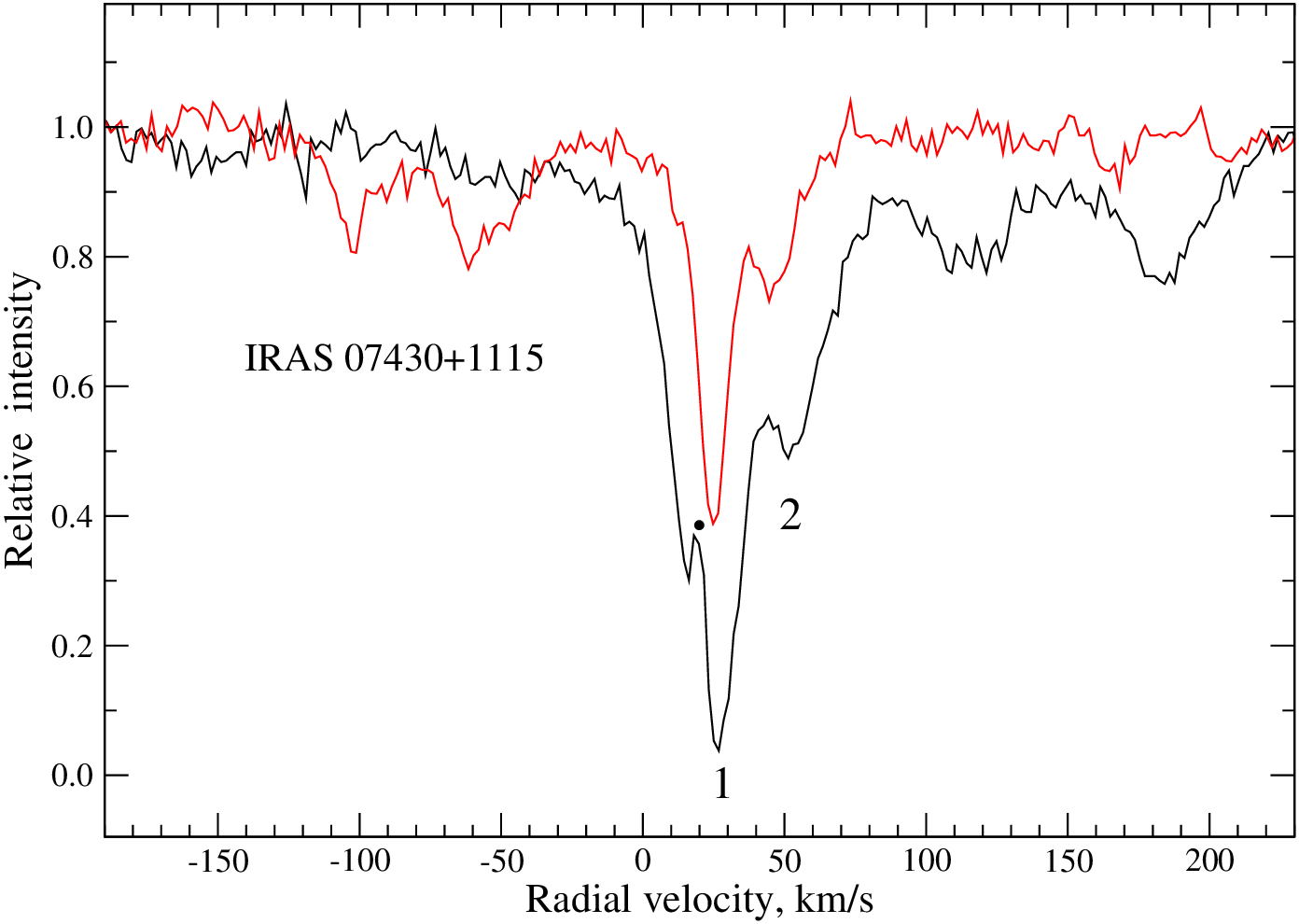}  
\includegraphics[angle=0,width=0.5\textwidth,bb=30 20 710 530,clip]{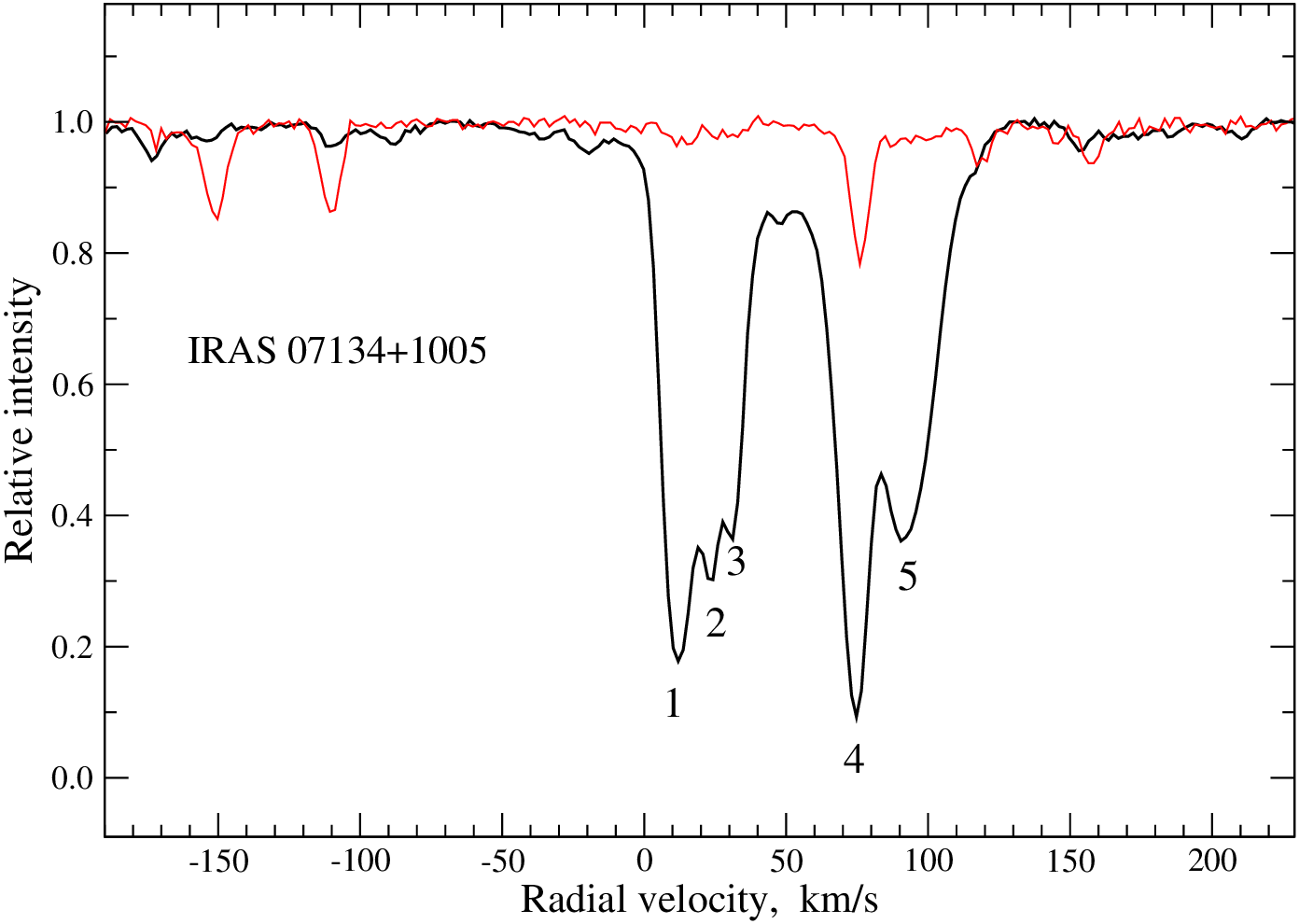}  
}
\caption{The NaI\,5890 (black line) and K\,I\,7699 (red line) absorption profiles in ``radial velocity--relative 
intensity'' coordinates. Panel (a) -- in the spectrum of IRAS\,07430, the dot indicates the position of
the telluric emission. Panels (b) --  the profiles of the same lines in the spectrum of the distant
post-AGB star IRAS\,07134. }
\label{KNa}
\end{figure}

\subsection{On Interstellar Features in the Spectra of C-Rich Stars}

Having the high-quality observations, we have identified diffuse interstellar bands (DIBs) in the spectra
of the studied post-AGB stars and measured their equivalent widths W$_{\lambda}$. This data allows, using
the published calibrations, to estimate the color excess E(B$-$V) and the interstellar absorption A$_V$,
the value of which is necessary in the problem of refining the distance and luminosity of the star. This
procedure is especially important for the star in the IRAS\,07430 system due to its paradoxically low
luminosity~[44]. First of all, let us consider the interstellar components of the
NaI D-line profiles in the IRAS\,07430 spectrum. The profile of one of the NaI D-lines is shown in Fig.\,7a
together with the K\,I\,7699\,\AA{} line. To clarify the velocity pattern in the IRAS\,07430 system, let us
compare the multicomponent profiles of the D2~NaI\,5890 line in the spectra of IRAS\,07430 and the distant
post-AGB star in  IRAS\,07134, which has galactic coordinates close to IRAS\,07430, shown in Fig.\,7.
As follows from the publication~[59], the positions of the three short-wave components 1--3 in the
spectra of IRAS\,07134 do not change with time within the measurement accuracy.
This  stability confirms their formation in the interstellar and circumstellar medium. The position of
the deep component ``4'' is consistent with the position of the Swan bands, which indicates its formation
in the circumstellar envelope. The most long-wavelength component ``5'' is photospheric: its temporal
behavior is consistent with the behavior of other photospheric absorptions in Table~3 from the paper~[59].

We have also performed a thorough search for the diffuse interstellar features (DIBs) in the spectra of
IRAS\,07430. However, only an intense interstellar band with broad wings at $\lambda$6281\,\AA{} was
reliably identified. The position of its core (Vr$\approx$23\,km/s) corresponds to a velocity close to
the one based on the position of the interstellar KI\,7699 line. However, due to the blending and noise
in the spectrum, the presence of other known DIBs, expectedly weak in the spectra of the star located
close to us, and moreover, rather remote from the galactic plane (the galactic longitude of IRAS\,07430
is b$>17\degr$) has not been confirmed.

\section{Conclusions}

In general, we can conclude that the kinematic state of the outflowing atmospheres of single C-rich
post-AGB stars is unstable. The effects that could be common for them are: variability of Vr
pattern due to pulsations, variability of complex H$\alpha$ profiles, the appearance of emission
in molecular bands, splitting of absorptions with a low lower-level excitation potential, as well
as velocity stratification in the atmosphere of the central star. It should be stressed that the
stars of the studied sample have effective temperatures in the range of log Teff$\approx3.7\div3.9$,
which is in good agreement with the temperatures of post-AGB stars with a possible photometric
instability $\Delta$V$>0.5^m$, caused by radial pulsations from Aikawa~[64]. In approximately
the same temperature range, theoretical modeling of radial pulsations in the atmospheres of post-AGB stars,
taking into account convection, was performed by Yu.A.~Fadeyev~[65].
In the spectra of C-rich stars of different luminosity (and, consequently, having different initial masses)
that have moved away from the AGB stage to different degrees, various combinations of the listed effects are
observed. For example, in the case of IRAS\,07430, the expected spectral features are absent: radial
velocity stratification in the stellar atmosphere, peculiarity of the H$\alpha$ profile, and splitting
of strong heavy metal absorptions, which we have found in the spectra of the closest kindred star in
the IRAS\,z02229 system. At the same time, the spectra of HD\,56126 in the IRAS\,07134 system show
the maximum pulsation amplitude, differential radial velocity shifts in the atmosphere, asymmetric absorption
profiles, and a significant variability of the complex H$\alpha$ profile. The spectrum of IRAS\,07134 from
March 30, 2024 has an anomalous H$\alpha$ profile, which absorption core is split into two components.
The short-wave H$\alpha$ absorption component is shifted by 60\,km/s relative to the systemic velocity.

A comparison of the photometric variability parameter $\Delta$V from Table~1 and the pulsation amplitude
$A_{Vr}$ from Table~3 indicates the absence of a correlation between these parameters.

As follows from the data in Table~3, the maximum amplitude of A$_{\rm Vr}$ was recorded for stars in the
IRAS\,07134 and IRAS\,19500 systems. Moreover, it is these two stars, having the maximum temperatures
Teff$\ge$7500\,K in the sample, that have moved away from the asymptotic giant branch the most. The
obtained information about the pulsation amplitude allows us to make a preliminary conclusion about
the influence of the star’s mass and the degree of its advancement from the AGB to the PN stage on the
level of atmospheric instability.

An unexpected result was obtained that yet requires an explanation: given a huge difference in the
luminosity of stars in the IRAS\,z02229 and IRAS\,07430 systems, the equivalent width W$_\lambda$
of the oxygen triplet OI(7774) in their spectra is almost the same: 0.99 and 0.91\,\AA{}, respectively.
The paradox in the luminosity estimates of the central star of the IR source IRAS\,07430 may be due to 
the uncertainty in the parallax of nearby cool stars with extended dust envelopes (see for details
Andriantsaralaza et al.~[66]).

\section*{ACKNOWLEDGMENTS}

Observations with the SAO RAS telescopes are supported by the Ministry of Science and Higher
Education of the Russian Federation. The renovation of telescope equipment is currently provided
within the national project ``Science and Universities''. The study made use of the SIMBAD, VALD,
SAO/NASA ADS, ASAS-SN and Gaia~DR3 astronomical databases.

\end{document}